\documentclass[aip,pop,
preprint,%
]{revtex4-1}

\usepackage{graphicx}
\usepackage{dcolumn}
\usepackage{bm}

\begin{document}

\preprint{}

\title{Plasma wakefield acceleration studies using the quasi-static code WAKE}%

\author{Neeraj Jain}
\thanks{Work done at Institute for Research in electronics and Applied Physics,
 University of Maryland, College Park, MD, USA}
 \affiliation{Max Planck Institute for Solar System Research,
Justus-von-Liebig-Weg 3, 37077, G\"ottingen, Germany}

\author{John Palastro}
\thanks{Work done at Institute for Research in electronics and Applied Physics,
 University of Maryland, College Park, MD, USA}
 \affiliation{Icarus Research Inc., P.O. Box 30780, Bethesda, Maryland 20824-0780}
 
\author{T. M. Antonsen Jr.}

\affiliation{Institute for Research in electronics and Applied Physics,
 University of Maryland, College Park, MD, USA}%
  \author{Warren B. Mori}
  \affiliation{University of California, Los Angeles, CA, USA}
  \author{Weiming An}
    \affiliation{University of California, Los Angeles, CA, USA}
\date{\today}

\begin{abstract}
The quasi-static code WAKE [P. Mora and T. Antonsen, Phys. Plasmas {\bf 4}, 217(1997)] is upgraded to model the propagation of an ultra-relativistic charged particle beam  through a warm background plasma in plasma wakefield acceleration. The upgraded code is benchmarked against the full particle-in-cell code OSIRIS [Hemker et al., Phys. Rev. ST Accel. Beams {\bf 3}, 061301(2000)] and the quasi-static code QuickPIC [Huang et al., J. Comp. Phys. {\bf 217}, 658 (2006)]. The effect of non-zero plasma temperature on the peak accelerating electric field is studied for a two bunch electron beam driver with parameters corresponding to the plasma wakefield acceleration experiments at FACET. It is shown that plasma temperature does not affect the energy gain and spread of the accelerated particles despite suppressing the peak accelerating electric field. The role of plasma temperature in improving the numerical convergence of the electric field with the grid resolution is discussed.  
\end{abstract}

\maketitle

\section{\label{introduction} Introduction}
In plasma wakefield acceleration (PWFA), a relativistic electron beam  propagates through a plasma and drives electric and magnetic fields known as wakefields \cite{chen1985}. In the blowout regime \cite{rosenzweig1991,lu2006}, a short (approximately one plasma wavelength $k_p^{-1}=c/\omega_{p}$ long, where $c$ is the speed of light and $\omega_p$ is the plasma frequency) and high current electron beam driver with density $n_b$ larger than the plasma density $n_p$ radially expels all the plasma electrons in its vicinity. As the beam passes by, the expelled electrons are pulled back towards the beam axis by the positive charge of the background plasma ions. The plasma electrons falling back on the axis generate a large longitudinal electric field. If a witness electron bunch is suitably placed in this wake, it can be accelerated to high energies through a transfer of energy from the drive beam to the witness beam via the plasma wakefields. This concept has been demonstrated by 
experiments in which electrons in the front of a 42 GeV electron beam created a wake which doubled the energy of electrons in the tail of the beam in only 85 cm of plasma \cite{hogan2000,blumenfeld2007}.

In order to better understand the physics in existing experiments and guide
future experiments, kinetic simulation codes with efficient algorithms are
required. In full particle-in-cell (PIC) models, such as in OSIRIS
\cite{hemker2000}, numerical stability conditions require resolving 
every scale thereby making full PIC models computationally expensive.
Computational efficiency can be achieved using either boosted frames \cite{martins2010} or the 
quasi-static approximation (QSA). In the QSA, the disparity of the time scales of the
evolution of beam driver and that of the background plasma allows one to achieve computational
efficiency. For highly relativistic electron beam drivers, the time scale of
evolution is the betatron period $\tau_b=\sqrt{2\gamma_b}\lambda_p/c$ which is
much larger than the plasma time scale $\lambda_p/c$ where $\lambda_p=2\pi/k_p$ is the plasma wavelength. Therefore in the QSA, the plasma
response is calculated on a fast time scale assuming a fixed beam driver. The
driver is then evolved over longer time scales. 

The codes QuickPIC [Huang et al., 2006]\cite{huang2006}, WAKE \cite{mora1997} and LCODE \cite{lotov2003} utilize the quasi-static approximation. The code WAKE was originally written for laser pulse propagation in a kinetic, cold and relativistic plasma. It is a 2D3V (two spatial -$r,z$ - and three velocity components) code which can handle both cylindrical and slab geometries. This code was upgraded to include the trapping of energetic plasma particles \cite{morshed2010}. The trapping was implemented in the code by promoting plasma particles satisfying threshold conditions to become 'beam particles' for which full equations of motion rather than quasi-static equations are solved. The code was then benchmarked against experiments, the full PIC code OSIRIS and the  three dimensional (3D) quasi-static code QuickPIC \cite{morshed2010}. The code was also benchmarked for PWFA studies for a non-evolving beam driver \cite{morshed2010}. 

In this paper we further upgrade the code to include the evolution of the electron beam driver as it propagates through the plasma. In the new version of the code, the background plasma can have non-zero temperature. The upgraded code is benchmarked against 3D OSIRIS and 3D QuickPIC simulations. The effect of plasma temperature on the amplitude of the accelerating electric field is studied for a two bunch electron beam driver with parameters representative of the PWFA experiments at FACET. Theoretical  studies using one-dimensional warm fluid theory show that non-zero plasma temperature limits the amplitude of the electric field  \cite{coffey1971,katsouleas1988}. On the other hand, one-dimensional (1D) Vlasov simulations show that the initial plasma temperature is reduced in the first accelerating bucket behind the driver and
thus the wake amplitude becomes insensitive to initial plasma temperature.
\cite{krall1991}. In the blowout regime in 2-D, the on-axis longitudinal electric field forms a sharp peak behind the beam driver. Two dimensional simulations using the quasi-static code LCODE have shown that the amplitude of the spike is suppressed for non-zero plasma temperature \cite{lotov2003}. Here, we show that although the non-zero plasma temperature reduces the amplitude of the peak in two dimensions (2D), in agreement with other studies \cite{lotov2003}, it does not affect the energy gain or spread of the accelerated particles.  Additionally, the peak electric field converges slowly with grid resolution for a cold plasma  \cite{lee2000}. We show that a non-zero plasma temperature can provide a faster numerical convergence for the electric field values.    

The next section presents the quasi-static equations for warm plasma particles and the full equations of motion for beam particles. Section \ref{sec:benchmarks} presents the benchmark studies against the codes OSIRIS and QuickPIC. The effect of plasma temperature is discussed in section \ref{sec:temperature}. Section \ref{sec:conclusion} ends the manuscript with our conclusions. 
\section{\label{sec:equations} Model}
\subsection{Equations for warm plasma and evolution of beam driver}
 We consider propagation of an ultra-relativistic cold electron beam through a
plasma with a non-zero initial temperature along the axis ($z$) of an azimuthally
symmetric ($\partial/\partial \theta = 0$) cylindrical coordinate system. Under
the quasi-static approximation, the plasma evolves on a time scale much faster than beam
evolution time scale. In WAKE, the equations of motion of plasma particles are
solved for a fixed beam current on a spatial computational domain that
follows the electron beam. The axial coordinate ($\xi$) in the moving computational domain
is  written as $\xi=ct-z$. Here positive values of $\xi$ represents the 
distance back from the head of the beam driver. 

In the transverse Coulomb gauge and azimuthal symmetry, the electromagnetic
fields are described by the electro-static potential $\phi$ and vector potential
$\mathbf{A}=(0,0,A_z)$. The equations of motion for
plasma particles can be obtained from the Hamiltonian $H(P_z,p_r,p_{\theta},r_p,\xi)=\gamma m_ec^2+q\phi$, where $\gamma=[1+(p_r^2+p_{\theta}^2+p_z^2)/m_e^2c^2]^{1/2}$, $P_z=p_z+qA_z/c$ , $\mathbf{p}=(p_r,p_{\theta},p_z)$ and $r_p$ are the relativistic factor, z-component of canonical momentum,  linear momentum and radial position of plasma particles, respectively. Under the quasi-static approximation, the Hamiltonian depends on $z$ and $t$ in the combination $\xi=ct-z$. This gives constancy of $H-cP_z=\gamma m_ec^2-cp_z+q\psi$, where $\psi=\phi-A_z$. The value of this constant of motion can be obtained from the unperturbed state of the plasma before the arrival of the beam driver. 
\begin{eqnarray}
 \gamma m_e c^2-cp_z+q\psi 
 &=& \gamma_0 m_e c^2-cp_{z0}.
 \label{eq:constant_motion}
\end{eqnarray}
Here $\gamma_0$ and $p_{z0}$ are the relativistic factor and z-component of linear momentum of a plasma particle due to the initial non-zero temperature. The  constant of motion for finite plasma temperature was earlier used to find trapping conditions of  plasma particles \cite{zeng2012}. Replacing  $\gamma$ in Eq. \ref{eq:constant_motion} by its expression in terms of $p_r$, $p_{\theta}$ and $p_z$, we find,
\begin{eqnarray}
  \gamma & = & \frac{1 + (p_r^2+p_{\theta}^2)/m_e^2c^2  + (\gamma_0
  - p_{z 0}/m_ec-q\psi/m_ec^2)^2}{2 (\gamma_0
  - p_{z 0}/m_ec-q\psi/m_ec^2)} \label{eq:gamma}\\
  p_z & = & \frac{1 + (p_r^2+p_{\theta}^2)/m_e^2c^2  - (\gamma_0
  - p_{z 0}/m_ec-q\psi/m_ec^2)^2}{2 (\gamma_0
  - p_{z 0}/m_ec-q\psi/m_ec^2)}\label{eq:pz}
  \end{eqnarray}
The azimuthal component of linear momentum can be written as $p_{\theta}=l_z/r_p$, where $l_z$ is the z-component of angular momentum of the plasma particles and is a constant of their motion. Using the transformation $\xi=ct-z$ and Eq. (\ref{eq:constant_motion}),the  radial components of the equations of motion of the plasma particles become
\begin{eqnarray}
 \frac{dp_r}{d\xi}&=&\frac{\gamma}{c(\gamma_0-p_{z0}/m_ec-q\psi/m_ec^2)}
 \left[-q\frac{\partial \psi}{\partial r}+\frac{l_z^2}{\gamma m_er_p^3}\right]
 +\frac{qB_{\theta}}{c}\\
 \frac{dr_p}{d\xi}&=&\frac{p_r}{m_ec(\gamma_0-p_{z0}/m_ec-q\psi/m_ec^2)}
\end{eqnarray}
where $B_{\theta}=-\partial A_z/\partial r$. The quasi-static equations for the plasma wakefields  can be obtained 
from Maxwell's equations using the transformation $\xi=ct-z$.
\begin{eqnarray}
 \frac{1}{r}\frac{\partial}{\partial r}(rB_{\theta})
 &=& \frac{\partial^2\psi}{\partial \xi^2}+\frac{4\pi}{c}J_z
 \label{eq:btheta}\\
\frac{\partial^2\psi}{\partial \xi \partial r} 
&=& \frac{4\pi}{c}J_r\label{eq:psi}
\end{eqnarray}
Here, the total current $\mathbf{J}=\mathbf{J}_{p}+\mathbf{J}_{b}$ has contributions both from plasma current $\mathbf{J}_p$ and beam current $\mathbf{J}_b$. 

Driver beam particles are evolved on a longer time scale according to the following equations of motion.
\begin{eqnarray}
 \frac{d\mathbf{p}_{b\perp}}{dt}&=&-q_b\nabla_{\perp}\psi-q_b(1-\frac{v_{bz}}{c})\nabla_{\perp}A_z\label{eq:beam_p_perp}\\
 \frac{dp_{bz}}{dt}&=&q_b\frac{\partial\psi}{\partial\xi}
 +q_b\frac{\mathbf{v}_{b\perp}}{c}\times[\nabla\times A_z\hat{z}]\label{eq:beam_pz}\\
 \frac{d\mathbf{x}_{b\perp}}{dt}&=&\frac{\mathbf{p}_{b\perp}}{m_b\gamma_b}\label{eq:beam_xperp}\\
 \frac{d\xi_b}{dt}&=&c-\frac{p_{bz}}{m_b\gamma_b}\label{eq:beam_z}
\end{eqnarray}
Here a suffix 'b' has been added to the beam particle variables in order to distinguish them from plasma particles. First, equations for the plasma particles and the wakefields, Eqs.(\ref{eq:gamma})-(\ref{eq:psi}), are solved in the $r$-$\xi$ computational domain and then the equations of motion for beam particles, Eqs. (\ref{eq:beam_p_perp})-(\ref{eq:beam_z}), are advanced in time $t$. 
\subsection{Two bunch electron beam driver}
An electron beam driver with two electron bunches, namely, a drive and a witness bunch, is loaded in the simulations. The two bunch driver corresponds to plasma wakefield experiments at the Facilities for Accelerator science and Experimental Test beams (FACET)\cite{hogan2010}.  The witness bunch follows the drive bunch and is accelerated in the wakefield generated by the drive bunch. The separation between the two electron bunches is chosen to optimize the quality of the accelerated bunch, i.e., to achieve high energy gain and low energy spread.  
The number density of the two bunch driver is expressed as,
\begin{eqnarray}
 n_b(r,\xi)&=&e^{-\frac{r^2}{2\sigma_r^2}}
 \left[
 n_{d}e^{-\frac{(\xi-\xi_d)^2}{2\sigma_{zd}^2}}
 + n_{w}e^{-\frac{(\xi-\xi_w)^2}{2\sigma_{zw}^2}}
 \right]
 \label{eq:two_bunch}
\end{eqnarray}
The drive and witness bunches are centered at ($0,\xi_d$) and ($0,\xi_w$) with peak densities $n_{d}$ and $n_{w}$, which fall off along the beam axis  in a distance of $\sqrt{2}\sigma_{zd}$ and $\sqrt{2}\sigma_{zw}$, respectively. The radial size of both  bunches is  $\sqrt{2}\sigma_r$.  A single electron bunch driver can be obtained by setting $n_{w}=0$ in Eq. (\ref{eq:two_bunch}). 
\section{\label{sec:benchmarks} Benchmark studies: Comparison with OSIRIS and QuickPIC}
We benchmark the code WAKE for plasma wakefield acceleration studies against
the full particle-in-cell simulation code OSIRIS \cite{hemker2000} and the 
3D quasi-static code Quick-PIC \cite{huang2006}. In the OSIRIS simulations of Hemker et
al. \cite{hemker2000}, a single bunch electron beam driver was used and thus
we take $n_{w}=0$. We take the $r$-$\xi$ computational domain size to be 
$10\,c/\omega_p\times 25\, c/\omega_p$ with $200\times 500$ grid points. The
cold background plasma of density $n_p=2.1\times10^{14}\,\mathrm{cm}^{-3}$
($c/\omega_p=0.367\,\mathrm{mm}$) is modeled using 9 particles per cell. The
peak density of the electron beam driver is $n_{d}=7.56\times
10^{14}\,\mathrm{cm}^{-3}=3.6\,n_p$ with $\sigma_r=70
\,\mu\mathrm{m}=0.19c/\omega_p$ and $\sigma_{zd}=0.63
\,\mathrm{mm}=1.72\,c/\omega_p$. The beam driver with 30 GeV initial energy is
modeled using a total of $2.5\times 10^6$ simulations particles giving an
average number of 25 particles per cell. These parameters are the same as chosen
by Hemker et al. \cite{hemker2000}.   

\begin{figure}
 \includegraphics[height=0.4\textheight,width=\textwidth]{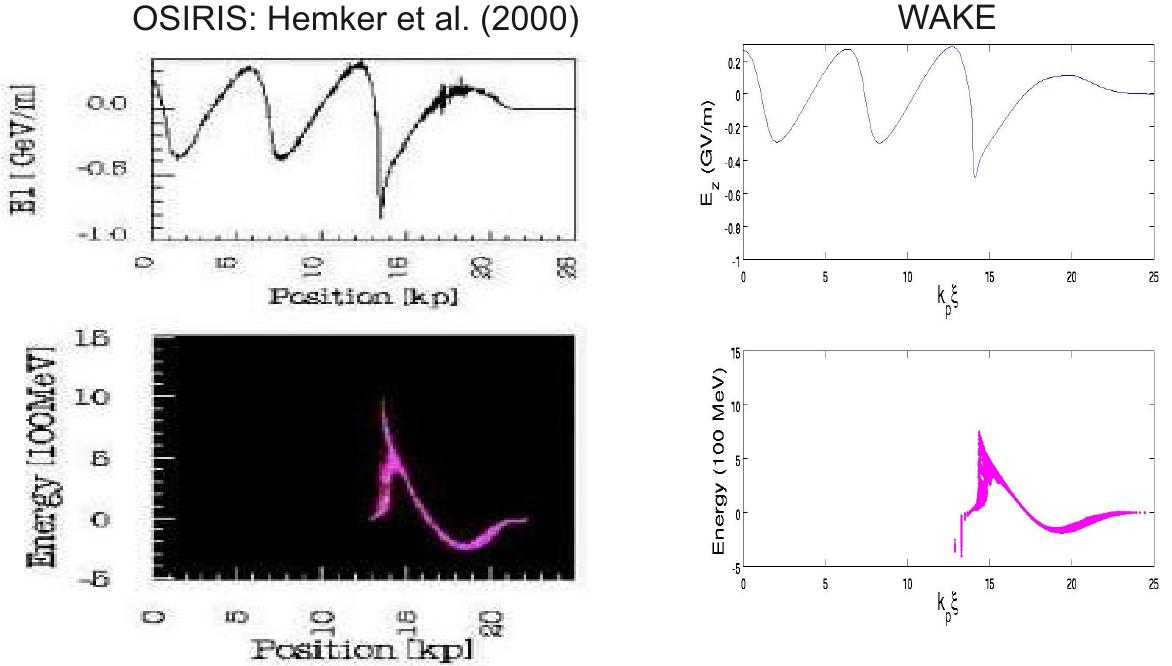}
 \caption{\label{fig:benchmark_OSIRIS}Comparison of the results of the OSIRIS simulations (left column) adopted from Fig. 1 of Hemker et al.\cite{hemker2000} and those of the WAKE simulations (right column). Line out of longitudinal  electric field $ E_z$ (top panel) along $r=0$ (axis of beam) and energy gain/loss of beam electrons (bottom panel) after
propagating a distance of 1.4 m. The beam
driver moves to the right and the color in the bottom panel represents
$\xi$-positions of beam electrons.}
\end{figure}

The line-out of the longitudinal  electric field $E_z$ on the axis of the beam driver after a propagation distance of 1.4 m is in very good agreement with the OSIRIS results (compare the left and right columns in Fig. \ref{fig:benchmark_OSIRIS}). However, there are mismatches at the negative peaks of $E_z$. The negative spikes behind the beam driver are quite sensitive to grid resolution and implementation of current deposition schemes in codes \cite{lee2000}. We shall revisit the issue of grid resolution dependent electric field spikes in later sections of this paper. The energy gain/loss of beam particles is also in good agreement with OSIRIS results. The maximum energy gain in OSIRIS simulations is higher than that in WAKE simulations due to the deeper electric field spikes in the OSIRIS simulations.  

\begin{center}
\begin{table}
\begin{tabular}{|lll|}
\hline
$n_p$ & &5.0$\times 10^{16}$ cm$^{-3}$\\
 $n_{d}$ & &1.78$\times 10^{17}$ cm$^{-3}$ (3.56 $n_p$)\\
  $n_{w}$ & &1.42$\times 10^{17}$ cm$^{-3}$ (2.85 $n_p$)\\
 $\sigma_r$ &  & 10 $\mu$m (0.42 $k_p^{-1}$) \\
 $\sigma_{zd}$ &  & 34.1 $\mu$m (1.44 $k_p^{-1}$) \\
  $\sigma_{zw}$ &  & 19.3 $\mu$m (0.81 $k_p^{-1}$) \\
   $\xi_w-\xi_d$ &  & 130 $\mu$m (5.48 $k_p^{-1}$) \\
   Beam energy & & 23 GeV\\
\hline
\end{tabular}
\caption{\label{tab:facet_parameters}Plasma and beam parameters
used for WAKE simulations of two bunch electron beam driver. Here $k_p=\omega_p/c$. These
parameters are similar to those in Quick-PIC simulations \cite{an2010} and
are representative of two bunch electron beam driver experiments at FACET.}
\end{table}
\end{center}

\begin{figure}
 \includegraphics[height=0.3\textheight,width=0.4\textwidth]
 {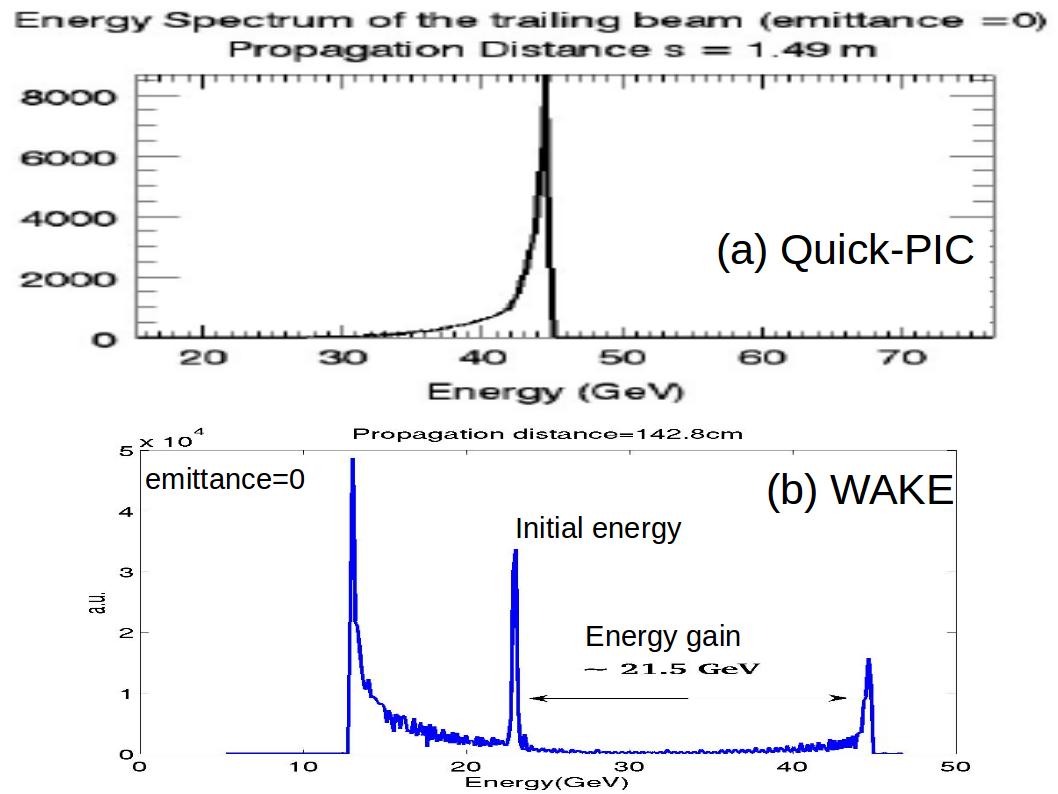}
 \caption{\label{fig:benchmark_QuickPIC}Energy spectrum of the witness bunch in
Quick-PIC simulations \cite{private_communication_an}(a) and of the two bunch beam driver in WAKE simulations
(b).}
\end{figure}

For the comparison of WAKE and Quick-PIC simulations,
we consider a two bunch electron beam driver. The plasma and beam parameters
for the simulations are shown in Table \ref{tab:facet_parameters}. The density
of the cold
background plasma in this case is taken to be $n_p=5\times
10^{16}\,\mathrm{cm}^{-3}$ giving $c/\omega_p=23.7 \mu\mathrm{m}$. These
parameters are similar to those chosen by
An et al. \cite{an2010} except that the initial emittance of the beam is
zero in our case. In both QuickPIC and WAKE simulations, the energy spectrum of
the witness bunch
 peaks at approximately 44.5 GeV after propagating  a distance of $\approx 1.49$
m (Fig. \ref{fig:benchmark_QuickPIC}). Since the initial energy of the witness
bunch is 23 GeV, the energy gain is approximately 21.5 GeV. Thus, the results of
QuickPIC and WAKE simulations are in good agreement. 

\section{\label{sec:temperature}Effect of plasma temperature on electric field
spike}
\begin{figure}
 \includegraphics[height=0.3\textheight,width=0.5\textwidth]
 {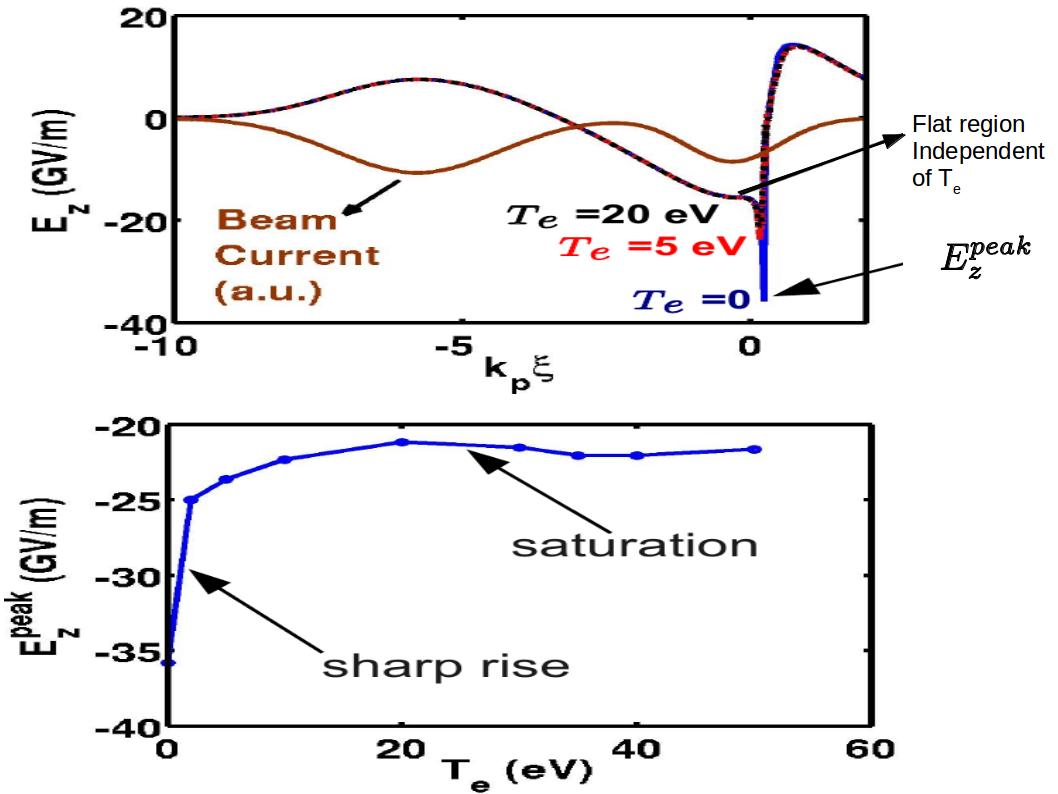}
 \caption{\label{fig:Ez_Te} Two bunch driver results: Profiles of $E_z$ for
different plasma temperatures and profile of the beam current along the
axis of the beam (top). Variation of the peak of the electric field spike with
plasma temperature (bottom).}
\end{figure}
It has earlier been reported that non-zero plasma temperature can suppress the
electric field peak that forms behind the beam driver \cite{lotov2003}. We
study, using WAKE, the effect of non-zero plasma temperature on the electric field spike
for the two electron bunch case. The plasma and beam parameters for the WAKE
 simulations correspond to two bunch experiments at FACET and are shown in
Table \ref{tab:facet_parameters}. 

Figure \ref{fig:Ez_Te} (top panel) shows profiles of the longitudinal electric
field $E_z$ along the axis of the beam for several values of plasma
temperature that would be expected in the experiments. It can be clearly seen that non-zero
plasma temperature does not affect the longitudinal electric field except at
the spike. Consistent with earlier studies \cite{lotov2003}, the magnitude of
the spike amplitude $E_z^{peak}$ drops when the temperature is non-zero. In the bottom panel, we show the 
variation of $E_z^{peak}$ with plasma temperature $T_e$. As the temperature
increases from zero to a small but finite value, $|E_z^{peak}|$ drops sharply.
The sharp drop is followed by a slow drop  until $T_e\approx 20$ eV at
which value $|E_z^{peak}|$ saturates. Essentially, the spike disappears.

\begin{figure}
 \includegraphics[height=0.3\textheight,width=0.4\textwidth]
 {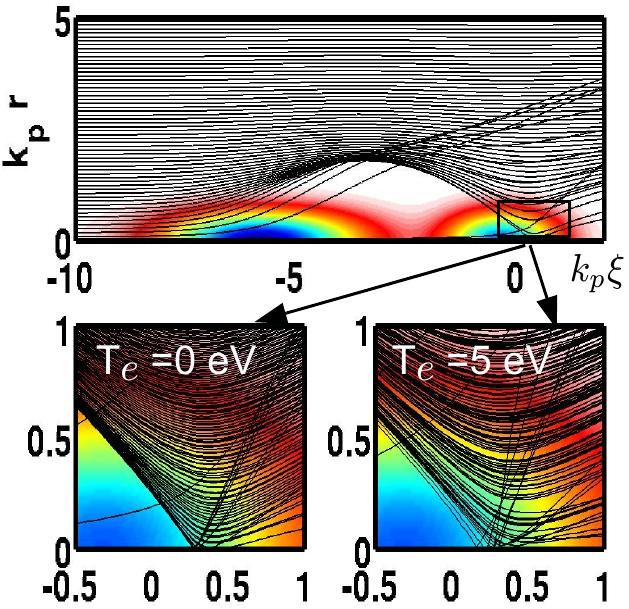}
 \caption{\label{fig:trajectories_twobunch}Trajectories (black lines) of
plasma electrons for a two bunch electron beam driver whose current
density is represented by colors (top). The bottom two panels show enlarged views
of the axis crossing of the trajectories behind the driver bunch for plasma
temperatures $T_e=0$ (left) and $T_e=5$ eV (right).}
\end{figure}

\begin{figure}
 \includegraphics[height=0.3\textheight,width=0.4\textwidth]
 {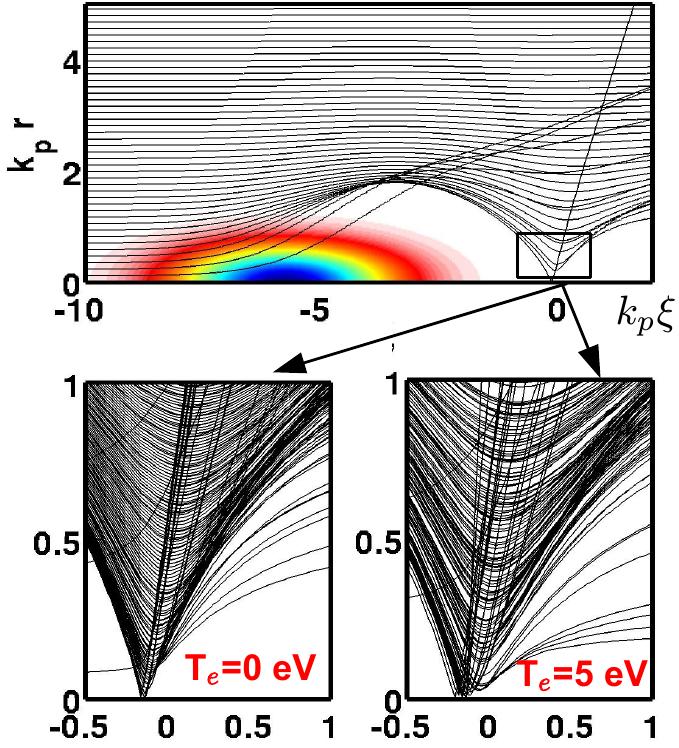}
 \caption{\label{fig:trajectories_singlebunch}Trajectories (black lines) of
plasma electrons for a single bunch electron beam driver whose current
density is represented by colors (top). The bottom two panels show enlarged view
of the axis crossing of the trajectories behind the beam driver for plasma
temperatures $T_e=0$ (left) and $T_e=5$ eV (right).}
\end{figure}

The drop in $|E_z^{peak}|$ with $T_e$ can be understood by looking at the
trajectories of plasma electrons, shown in Fig.
\ref{fig:trajectories_twobunch}. The plasma electrons expelled
radially outward by the beam driver fall back and cross the axis behind the
driver. The spike of the longitudinal electric field forms where these
electrons cross the axis. It can be seen in the bottom two panels of Fig.
\ref{fig:trajectories_twobunch} that warm electrons cross the axis
over a region that is broader than the one in the case of cold electrons.
This is because warm electrons have a distribution of initial velocities. 
The effect of plasma temperature on the trajectories of plasma electrons can be
more clearly seen in case of a single bunch driver. Figure
\ref{fig:trajectories_singlebunch} shows the trajectories of plasma electrons
for the case of a single bunch driver obtained by setting $n_{w}=0$ in the list of beam parameters shown in
Table \ref{tab:facet_parameters}. The other parameters are the same as for the
two bunch driver. For a single bunch driver, cold electrons cross the axis in a
much narrower region as compared with the two bunch driver case.
Again, non-zero plasma temperature spreads the trajectories of plasma electrons
over a broader region on the axis. This makes the charge density smaller in the
axis crossing region in the case of warm electrons, and thus, suppresses the electric
field spike. 

In the top panel of Fig. \ref{fig:Ez_Te}, the electric field profile has a flat
region (indicated by an arrow) near the spike due to the beam load \cite{tzoufras2008} of the witness
bunch behind the driver bunch. The flat region has relatively uniform
accelerating electric field and thus improves the quality of the accelerated
electrons by reducing the energy spread. The parameters of the driver and
witness bunch and distance between them can be tuned to optimize the flat
region (magnitude of electric field and extent of the region) for a high
quality accelerated beam. Although a major portion of the witness bunch sits
and is accelerated in the flat region, a finite
number of electrons (at and very close to the location of the spike) can be
accelerated by the spike electric field. Since the magnitude of the spike
electric field drops with increasing
temperature, a natural question arises: How does the plasma temperature affect
the energy gain and energy spread?

\begin{figure}
 \includegraphics[height=0.3\textheight,width=0.5\textwidth]
 {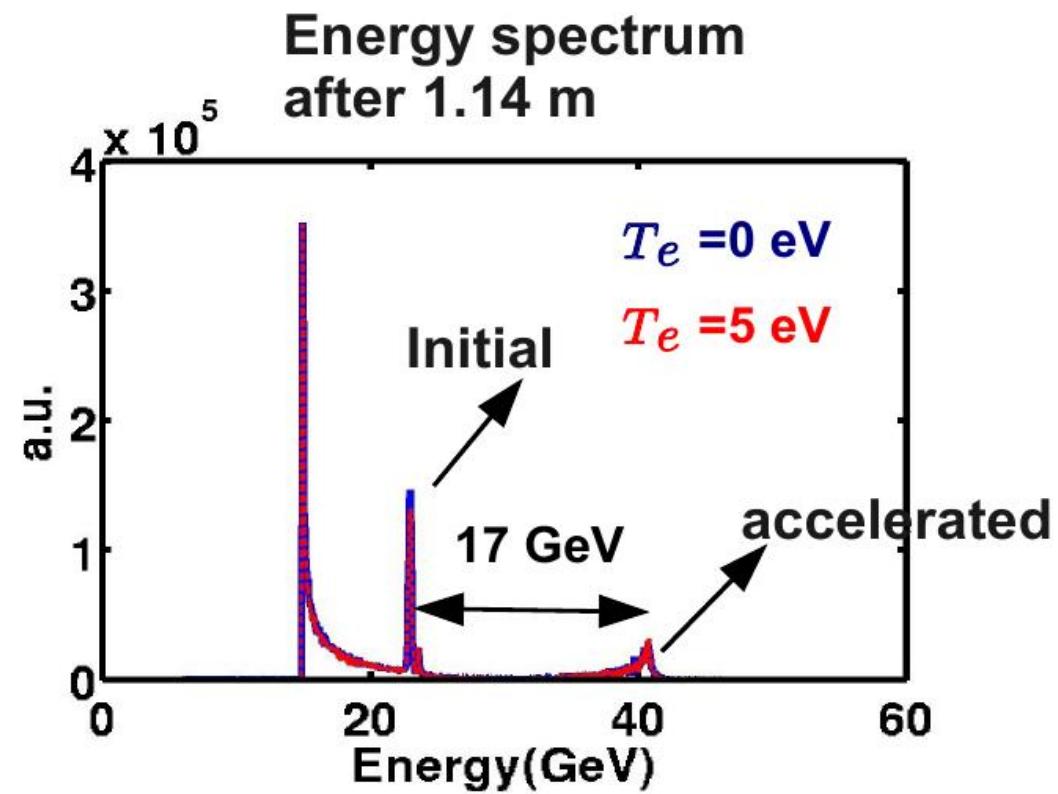}
 \caption{\label{fig:energy_spectrum_0ev_5ev} Two bunch driver results: Energy
spectrum of the beam electrons after propagation of a distance of 1.14 m for
$T_e=0$ and $T_e=5$ eV}
\end{figure}

\begin{figure}
 \includegraphics[height=0.4\textheight,width=0.5\textwidth]{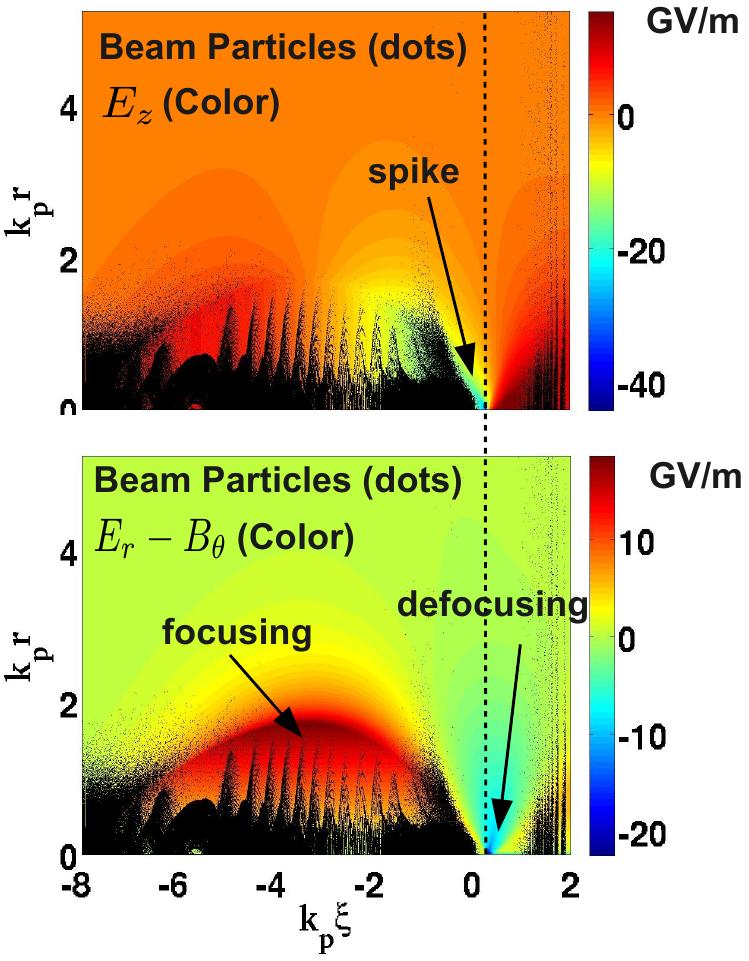}
 \caption{\label{fig:Ez_ErMinusBtheta_5ev} Two bunch driver results: Longitudinal 
electric field $E_z$ (top) and radial electric field $E_r-B_{\theta}$ (bottom)
in $r-\xi$ space. Black dots represent beam
electrons. A vertical dashed line from top to bottom shows the alignment of the
locations of electric field spike and defocusing regions.}
\end{figure}

Figure \ref{fig:energy_spectrum_0ev_5ev} shows the energy
spectrum of beam electrons after  1.14 meters of propagation for $T_e=0$ and
$T_e=5$ eV.  Based on the relatively  uniform electric field $\approx 15$ GV/m
in the
flat region in Fig. \ref{fig:Ez_Te}, the expected energy gain after 1.14 m of
propagation is $15 \times 1.14\approx 17$ GeV which is the same energy gain as obtained from
WAKE simulation and shown in Fig. \ref{fig:energy_spectrum_0ev_5ev}. The
expected maximum energy gains by the peaks of the electric field spikes for
$T_e=0$ and $T_e=5$ eV are $36 \times 1.14=41$ Gev and $24\times 1.14\approx
27.4$ GeV, respectively. However we do not see any such energy gains in Fig.
\ref{fig:energy_spectrum_0ev_5ev}. There is little difference
in the two energy
spectra. The reason for this is explained in Fig.
\ref{fig:Ez_ErMinusBtheta_5ev}. The location of the spike of the
longitudinal electric
field is indicated by an arrow in the top panel of Fig.
\ref{fig:Ez_ErMinusBtheta_5ev}. A dashed vertical line connecting the top and
bottom panel shows that the radial electric field in the bottom panel is inward
(towards the axis) and thus defocussing for electrons  at the location of the
 field spike. This defocussing radial electric field expels the beam
electrons at the location of the spike away from the axis. In Fig.
\ref{fig:Ez_ErMinusBtheta_5ev}, there are no beam electrons in the defocussing
region, and therefore, the spike of the longitudinal electric field does not contribute to
energy gain.

\begin{figure}
 \includegraphics[height=0.3\textheight,width=0.5\textwidth]
 {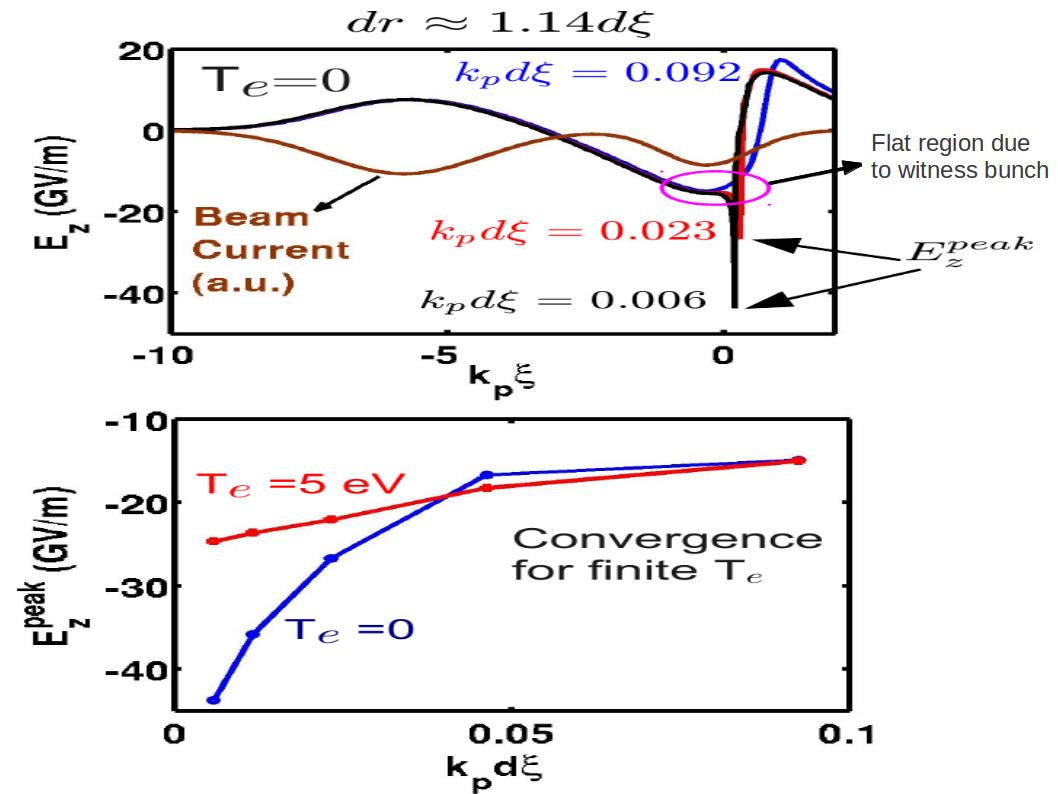}
 \caption{\label{fig:Ez_twobunch} Results for two bunch driver.
Profiles of the longitudinal electric field $E_z$
along the axis of the beam propagation shown for different grid
sizes when plasma electrons are cold (top). In the top panel,
profile of initial beam current along the beam axis is also shown for the reference. The peak
of $E_z$-spike as a
function of grid size $k_pd\xi$ for cold ($T_e=0$) and warm ($T_e=5$ eV)
electrons (bottom).}
\end{figure}

\begin{figure}
 \includegraphics[height=0.3\textheight,width=0.5\textwidth]
 {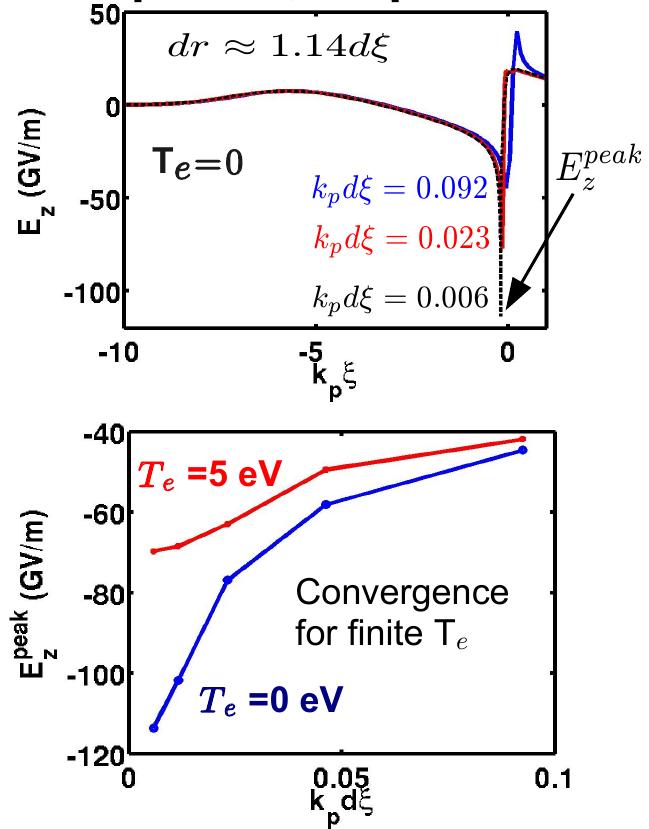}
 \caption{\label{fig:Ez_singlebunch} Results for single bunch driver.
Profiles of the longitudinal electric field $E_z$
along the axis of the beam propagation shown for different grid
sizes when plasma electrons are cold (top). The peak of $E_z$-spike as a
function of grid size $k_pd\xi$ for cold ($T_e=0$) and warm ($T_e=5$ eV)
electrons (bottom).}
\end{figure}

Although plasma temperature does not affect the energy gain and spread, it can effectively 
improve the numerical convergence of the electric field spike with grid size.
The electric field spike has been shown to depend on the grid
resolution in full PIC simulations using OSIRIS
\cite{lee2000}. The amplitude of the spike increases as grid resolution is
increased. We saw similar behavior of the spike in WAKE simulations,
as shown in Fig. \ref{fig:Ez_twobunch}.   The flat region in which the witness
bunch
sits does not depend on the
grid size, however the peak of the electric field spike does depend on grid size. The bottom panel in
Fig.
\ref{fig:Ez_twobunch} shows the dependence of  $E_z^{peak}$ on the grid size
$d\xi$ for $T_e=0$ and $T_e=5$ eV. For cold
plasma, the value of $E_z^{peak}$ does not seem to converge while it
tends to converge to a finite value for $T_e=5$ eV. Similar behavior is
observed for the case of a single bunch driver ($n_{w}=0$), shown in Fig.
\ref{fig:Ez_singlebunch}.  The reason for this behavior can be
understood as follows. As long as the grid size is
larger than the axis crossing region (confining charge) behind the electron beam
driver, reducing the grid size will increase the charge density, and thus,
the electric field at the spike. This is because the reduce grid volume still
contain the same amount of charge. Once the grid size is comparable to
the extent of the axis crossing region, the amplitude of the spike should
converge to a finite value.
For warm plasma, electrons cross the axis in a
relatively broader region as compared to the cold plasma. This makes
it possible to resolve the axis crossing region with a
relatively large grid size.

\section{\label{sec:conclusion}Conclusion}
We upgraded the quasi-static code WAKE to include the capabilities of modeling  the propagation of an electron (charged particle) beam driver through a warm background plasma in plasma wakefield acceleration. The code was benchmarked against (1) published 3D results from the full particle-in-cell code OSIRIS for a single bunch electron beam driver and (2) the 3D quasi-static code QuickPIC for two bunch electron scheme with parameters corresponding to experiments at FACET. For the two bunch scheme, the spike of the electric field which forms behind the driver bunch is suppressed for a range of  values of the plasma temperature attainable in plasma wakefield experiments. This is because non-zero plasma temperature leads to the axis crossing of the plasma electrons over a broader region, decreasing the charge density and thus electric field of the spike. However, the suppression of spike does not affect the energy gain and energy spread of the accelerated electrons because the spike is co-located with a 
defocussing region. Due to the broadening  of the axis crossing 
region of plasma electrons for  non-zero plasma temperature, the electric field spike  can be resolved with coarser grid as compared to the one required for cold plasma, thus improving the numerical convergence of the electric field spike with grid resolution.  

\begin{acknowledgments}
This work was supported by DoE grants DE SC0008491, DE SC0008316, DE FG02-92
ER40727, NSF ACI 1339893, ONR N000140911190 and DoE DESC0007970.
\end{acknowledgments}


\nocite{*}
%

\end{document}